\documentstyle[12pt]{article}
\title{Canonical Approach to String Theory in Massive
Background Fields}
\author{I.L. Buchbinder\\
\it Department of Theoretical Physics, \\
\it Tomsk State Pedagogical University, \\
\it Tomsk 634041, Russia
\and
V.D. Pershin\\
\it Department of Theoretical Physics, \\
\it Tomsk State University,\\
\it Tomsk 634050, Russia
\and
G.B. Toder\\
\it Department of Physics and Chemistry, \\
\it Omsk State Academy of Railway Communications,\\
\it Omsk 644046, Russia}
\date{}
\begin{document}
\begin{titlepage}
\maketitle
\thispagestyle{empty}

\begin{abstract}
A method of constructing a canonical gauge invariant quantum
formulation for a non-gauge classical theory depending on a
set of parameters is advanced
and then applied to the theory of closed bosonic string
interacting with  massive background fields.
Choosing an ordering prescription and developing a suitable
regularization technique we calculate quantum guage algebra
up to linear order in background fields.
Requirement of closure for the algebra leads
to equations of motion for massive background fields which appear to
be consistent with the structure of string spectrum.
\end{abstract}

\vfill

03.70.+k, 11.10.Ef, 11.17.+y

\end{titlepage}

\section{Introduction}

$\sigma-$model approach still remains an important method of
string theory investigation allowing one to derive string
amplitudes as expectation values of corresponding vertex
operators~\cite{callan} (see also the reviews~\cite{tseyt}).
The main result received within this approach by
means of perturbative path integral methods is that quantum
conformal invariance condition leads to effective equations of
motion for massless background fields.
This condition appears as independence of quantum effective
action on the conformal factor of two-dimensional metric or as
vanishing of renormalized operator of the energy-momentum trace.

It should be noted that according to the prescription \cite{Pol}
generally accepted in functional approaches to string theory
dynamical variables are  treated in different ways. Namely,
functional integration is carried out only over string
coordinates $X^\mu(\tau,\sigma)$ while components of
two-dimensional metric $g_{ab}(\tau,\sigma)$ are considered as
external fields. Then one demands the result of such an
integration to be independent on the conformal factor and the
integrand over $g_{ab}(\tau,\sigma)$ reduces to finite
dimensional integration over parameters specifying string world
sheet topologies. This prescription differs from the standard
field theory rules when functional integral is calculated over
every variable independently \footnote{In string theory this
independent integration would lead to appearance at the quantum
level of an extra degree of freedom connected with
two-dimensional gravity \cite{buchshap}.}.

This approach can be as well applied to string theory
interacting with massive background fields which is not
classically conformal invariant \cite{ovrut}. One demands that
operator of the energy-momentum vanish no matter whether the
corresponding classical action is conformal invariant or not. As
was shown in \cite{buchper,krykh} it gives rise to effective
equations of motion for massive background fields (See also
\cite{sathiap,ellw}).  Unfortunately, in the case of closed
string theory covariant approaches did not reproduce the full
set of correct linear equations of motion for massive background
fields.  Namely, the tracelessness condition on massive tensor
fields has been obtained neither in standard covariant
perturbation approaches~\cite{buchper} nor within the formalism
of exact renormalization group~\cite{ellw}.  So there exists a
problem of deriving the correct equations for massive
background fields in framework of $\sigma-$model approach.

Moreover, from general point of view the requirement of quantum
conformal invariance of string theory with massive background
fields means that a non-gauge classical theory depending on a
set of parameters is used for constructing of a quantum theory
that is gauge invariant under some special values of the
parameters.  Such a situation occurs in string theory if
interaction with massless dilaton, tachyon or any other massive
field is turned on and it leads to a general problem in the
theory of quantization.

The most natural and consistent approach to construction of
quantum models in modern theoretical physics is
the procedure of canonical quantization.
We consider this approach to be the
only completely consistent method for constructing quantum
theories and so every step of any quantization procedure
should be justified by an appropriate prescription within
canonical formulation.  So the general problem arising from
string theories is how to describe in terms of canonical
quantization construction of gauge invariant quantum theory
starting with a classical theory without this invariance.

The most general realization of this
procedure ensuring unitarity at the quantum level and
consistency of theory symmetries and dynamics is BFV
method~\cite{BFV,BatFr,Hen}. Due to this method
one should construct hamiltonian formulation of classical theory,
find out all first class constraints and calculate algebra of
their Poisson brackets.  Then one defines fermionic functional
$\Omega$ generating algebra of gauge transformations and bosonic
functional $H$ containing information of theory dynamics.
Quantum theory is consistent provided that the operator $\hat\Omega$
is nilpotent and conserved in time. The corresponding analysis
was carried out in \cite{hwang,fujiw} for free bosonic string
and in \cite{buch91,buch95} for bosonic string coupled to
massless background fields.

In the case of string theory interacting with massive background
fields components of two-dimensional metric should be treated
as external fields, otherwise classical equations of
motion would be inconsistent. As a consequence, classical
gauge symmetries are absent and it is impossible to construct
classical gauge functional $\Omega$. In this paper we propose a
prescription allowing for some models to construct quantum
operator $\hat\Omega$ starting with a classical theory without
first class constraints. Quantum theory is gauge invariant if
there exist values of theory parameters providing nilpotency and
conservation of operator $\hat\Omega$. Then to illustrate how
the prescription works we apply it to the theory of closed
bosonic string coupled to massive background fields and show
that correct linear equations of motion are produced.

Calculation of quantum gauge algebra for the string interacting
with background fields is not trivial for two reasons.
First of all, the ordering prescription usually used in free
string theory when all the ambiguity reduces to two constants
$\alpha(0)$, $\beta(0)$~\cite{green} is not enough in the case of string
interacting with background fields. To deal with this issue we
use a specific operators ordering with an ambiguity that was
absent in the free string theory but is relevant in our case.
Second, in presence of background fields quantum deformation
of the conformal algebra contains divergencies which should be
regularized in an appropriate way.  In order to solve this
problem we propose a kind of analytical regularization which
depends on a single parameter and ensures finiteness of all
contractions of the fundamental operators.

The paper is organized as follows. In Sec.~2 we describe a
general method of constructing a gauge invariant
canonical quantum formulation for a non-gauge classical theory.
Sec.~3 contains description of the theory of closed bosonic
string coupled to background fields of tachyon and of the first
massive level. We build hamiltonian formulation for the model
and show that it can be studied using the general method of the
previous section.  Sec.~4 is devoted to calculation of conformal
algebra.
Conclusion contains summary and outlooks.

\section{From a non-gauge classical theory to a gauge invariant
quantum theory}

Consider a system described by a hamiltonian
\begin{equation} H=H_{0}(a)+\lambda^{\alpha}T_{\alpha}(a)
\label{Hgen}
\end{equation}
where $H_0(a)=H_0(q,p,a)$, $T_{\alpha}(a)=T_{\alpha}(q,p,a)$ and
$q$, $p$ are canonically conjugated dynamical variables;
$a=a_i$ and $\lambda^{\alpha}$ are external parameters of the
theory.

We suppose that $T_{\alpha}(a)$ are some functions of the form
\begin{equation}
T_{\alpha}(a)=T^{(0)}_{\alpha}(a)+T^{(1)}_{\alpha}(a)
\end{equation}
and closed algebra in terms of Poisson brackets is formed by
$T^{(0)}_{\alpha}(a)$, not by $T_{\alpha}(a)$:
\begin{eqnarray}
\{ T^{(0)}_{\alpha}(a),T^{(0)}_{\beta}(a) \} & = &
T^{(0)}_{\gamma}(a)U^{\gamma} _{\alpha\beta}(a)
\nonumber
\\
\{ H_0(a),T^{(0)}_{\alpha}(a) \} & = &
T^{(0)}_{\gamma}(a)V^{\gamma} _{\alpha}(a)
\end{eqnarray}
Such a situation may occur, for example, if
$T^{(0)}_{\alpha}(a)$ correspond to a free gauge invariant
theory and $T^{(1)}_{\alpha}(a)$ describe a perturbation
spoiling gauge invariance.

At the quantum level both the algebras of
$T^{(0)}_{\alpha}(a)$ and $T_{\alpha}(a)$ are not closed in
general case
\begin{eqnarray}
[T_{\alpha}(a), T_{\beta}(a)] &=& i\hbar \bigl(
T_{\gamma}(a) U^{\gamma} _{\alpha\beta}(a) + A_{\alpha\beta}(a)
\bigr),
\nonumber
\\
{} [H_0(a),T_{\alpha}(a)]& = & i\hbar \bigl(
T_{\gamma}(a)V^{\gamma}_{\alpha}(a) + A_{\alpha}(a) \bigr),
\label{A}
\end{eqnarray}
and operators $A_{\alpha\beta}$, $A_\alpha$ do not vanish in the
limit $\hbar\rightarrow 0$ due to absence of classical gauge
invariance.

The main step of our approach consist in defining
of the quantum operators $\Omega$ and  $H$ as if the functions
$T_{\alpha}(a)$ were first class constraints:

\begin{eqnarray}
\Omega & = &
c^{\alpha}T_{\alpha}(a)-{1\over2}U^{\gamma}_{\alpha\beta}(a)
:{\cal P_{\gamma}} c^{\alpha} c^{\beta}:
\nonumber
\\
H & = & H_0 (a) + V^{\gamma}_{\alpha}(a) :{\cal P_{\gamma}}
c^{\alpha}:
\end{eqnarray}
where $:\quad:$ stands for some ordering of ghost fields. Square
of such an operator $\Omega$ and its time derivative take the
form
\begin{eqnarray}
\Omega^{2} & = & {1\over2}i\hbar
\bigl( A_{\alpha\beta}(a) + G_{\alpha\beta}(a) \bigr)
:c^{\alpha}c^{\beta}:
\nonumber \\
\frac{d\Omega}{dt}& = &
\frac{\partial\Omega}{\partial t}+[H,\Omega] = {}
\nonumber \\
& = & \left( \frac{\partial T_{\alpha}(a)}{\partial t}-
A_{\alpha}(a) - G_{\alpha}(a)\right) c^{\alpha}-{1\over2}
\frac{\partial U^{\gamma}_{\alpha\beta}(a)}{\partial t}
:{\cal P_{\gamma}}c^{\alpha}c^{\beta}:
\label{Omega}
\end{eqnarray}
where $G_{\alpha\beta}(a)$, $G_\alpha(a)$ are possible quantum
contributions of ghosts.

It is natural to suppose that every operator of the theory can
be decomposed in a linear combination of an irreducible set of
independent operators $O_M(q,p)$:
\begin{eqnarray}
& &A_{\alpha\beta}(a)+G_{\alpha\beta}(a) =
E^{M}_{\alpha\beta}(a)O_{M}(q,p) {,}
\nonumber \\
& &\frac{\partial T_{\alpha}(a)}{\partial t} - A_{\alpha}(a) -
G_{\alpha}(a) = E^{M}_{\alpha}(a)O_{M}(q,p) {,}
\end{eqnarray}
$E^{M}_{\alpha\beta}(a)$, $E^M_\alpha(a)$ being some $c-$valued
functions of the parameters $a$.

In general case $\Omega^2 \neq 0$ and $d\Omega/dt \neq 0$.
However, if equations
\begin{equation}
E^{M}_{\alpha\beta}(a) = 0, \quad
E^{M}_{\alpha}(a) = 0, \quad
\frac{\partial U^{\gamma}_{\alpha\beta}(a)}{\partial t}=0,
\label{a_0}
\end{equation}
have some solutions $a_i=a_i^{(0)}$ then operator $\Omega$ is
nilpotent and conserved for these specific values of parameters
and hence the quantum theory is gauge invariant. Thus, there
exists a possibility to construct quantum theory with given gauge
invariance that is absent at the classical level.

It is important that eqs.(\ref{a_0}) are not anomaly
cancellation conditions because an anomaly represents breaking
of classical symmetries at the quantum level. In the theory
under consideration classical symmetries are absent and
eqs.(\ref{a_0}) express conditions of quantum symmetries
existence.

In specific models eqs.(\ref{a_0}) may have no solutions at all
or, conversely, be fulfilled identically. An example of the
latter possibility was described in \cite{Fubini}.

\section{String in massive background fields}

As  an example where the described procedure really
works and leads to eqs.(\ref{a_0}) with non-trivial solutions
for parameters $a$ we consider closed bosonic string theory
coupled with background fields of tachyon and of the first
massive level. We will restrict ourselves by linear
approximation in background fields.~\footnote{An adequate
treatment of non-linear (interaction) terms is known to demand
non-perturbative methods \cite{Das}. Our aim here is just to
illustrate possibilities of the general prescription given in
the previous section.} It will be enough to establish
consistency of background fields dynamics with structure of the
corresponding massive level in string spectrum.

The theory is described by the classical action
\begin{eqnarray}
S&=&{}-{1\over 2\pi\alpha'} \int d^{2}\sigma\,
  \sqrt{-g} \bigl\{{1\over 2} g^{ab}\partial_{a} X^{\mu} \partial_{b}
  X^{\nu} \eta_{\mu\nu} + Q(X)
\nonumber \\
& & \qquad {}+  g^{ab} g^{cd} \partial_a X^\mu \partial_b X^\nu
          \partial_c X^\lambda \partial_d X^\kappa
        F^1_{\mu\nu\lambda\kappa}(X)
\nonumber \\
& &\qquad{}+ g^{ab}\varepsilon^{cd} \partial_a X^\mu \partial_b X^\nu
          \partial_c X^\lambda \partial_d X^\kappa
      F^2_{\mu\nu\lambda\kappa}(X)
\nonumber \\
& &\qquad {}+ \alpha' R^{(2)} g^{ab} \partial_a X^\mu \partial_b X^\nu
      W^1_{\mu\nu}(X) +
  \alpha' R^{(2)} \varepsilon^{ab} \partial_a X^\mu \partial_b X^\nu
      W^2_{\mu\nu}(X)
\nonumber \\
& &\qquad {}+  \alpha'^2 R^{(2)} R^{(2)} C(X) \bigr\},
\label{S}
\end{eqnarray}
$\sigma^a=(\tau,\sigma)$ are coordinates on string world sheet,
$R^{(2)}$ is scalar curvature of two-dimensional  metric
$g^{ab}$, $\eta_{\mu\nu}$ is Minkowski metric of
$D-$dimensional spacetime,  $Q(X)$ is tachyonic field and $F$,
$W$, $C$ are background fields of the first massive level in
string spectrum.  As was shown in \cite{buchper} all other
possible terms with four two-dimensional derivatives in
classical action are not essential and string interacts with
background fields of the first massive level only by means of
the terms presented in (\ref{S}).

If components of two-dimensional metric
$g_{ab}$ were considered as independent dynamical variables then
the corresponding classical equations of motion would be
fulfilled only for vanishing background fields:
\begin{eqnarray}
\frac{2\pi}{\sqrt{-g}} g_{ab} \frac{\delta S}{\delta g_{ab}}&=&
{} - (1/\alpha')\sqrt{-g} Q(X)
\nonumber
\\& &
{} + g^{ab} g^{cd} \partial_a x^\mu \partial_b x^\nu
\partial_c x^\lambda \partial_d x^\kappa (
- (1/\alpha') F^1_{\mu\nu\lambda\kappa}
+ \partial_\mu \partial_\nu W^1_{\lambda\kappa} )
\nonumber
\\& &
{}+   g^{ab} \varepsilon^{cd} \partial_a x^\mu
\partial_b x^\nu \partial_c x^\lambda \partial_d x^\kappa
( - (1/\alpha') F^2_{\mu\nu\lambda\kappa}
+ \partial_\mu \partial_\nu W^2_{\lambda\kappa} )
\nonumber
\\& &
{} +   g^{ab} D^2 x^\mu \partial_a x^\nu \partial_b x^\lambda
\partial_\mu W^1_{\nu\lambda}
{}+   g^{ac} g^{bd} D_a \partial_b x^\mu \partial_c x^\nu
\partial_d x^\lambda
 4 \partial_\lambda W^1_{\mu\nu}
\nonumber
\\& &
{}+   g^{ac} \varepsilon^{bd} D_a \partial_b x^\mu
\partial_c x^{[\nu} \partial_d x^{\lambda]}
2( \partial_\mu W^2_{\nu\lambda}
+  \partial_\nu W^2_{\mu\lambda}
-  \partial_\lambda W^2_{\mu\nu})
\nonumber
\\& &
{}+   g^{ac} \varepsilon^{bd} D_a \partial_b x^\mu
\partial_c x^{(\nu} \partial_d x^{\lambda)}
2( \partial_\nu W^2_{\mu\lambda}+\partial_\lambda W^2_{\mu\nu})
\nonumber
\\ & &
{}+   g^{ac} g^{bd} D_a \partial_b x^\mu D_c \partial_d x^\nu
 2 W^1_{\mu\nu}
{}+   g^{ac} \varepsilon^{bd} D_a \partial_b x^\mu
D_c \partial_d x^\nu
2 W^2_{\mu\nu}
\nonumber
\\& &
{}+   g^{ab} D_a D^2 x^\mu \partial_b x^\nu
2 W^1_{\mu\nu}
{}+   \varepsilon^{ab} D_a D^2 x^\mu \partial_b x^\nu
2 W^2_{\mu\nu} +{}
\nonumber
\\& &
{}+   R^{(2)} g^{ab} \partial_a x^\mu \partial_b x^\nu
2 (- W^1_{\mu\nu} +  \alpha'\partial_\mu \partial_\nu C)
\nonumber
\\& &
{}+   R^{(2)} \varepsilon^{ab} \partial_a x^\mu \partial_b x^\nu
(- 2 W^2_{\mu\nu})
{}+   g^{ab} \partial_a x^\mu \partial_b R^{(2)}
  4 \alpha'\partial_\mu \, C
\nonumber
\\& &
{}+   R^{(2)} D^2 x^\mu
2 \alpha'\partial_\mu C
{}+   D^2 R^{(2)} \, 2\alpha' C
+   R^{(2)} R^{(2)} (- \alpha'C) = 0 {.}
\end{eqnarray}
Analogous situation arises in the theory of string interacting
with massless dilaton field~\cite{buch91}. Therefore to
construct the theory with non-trivial massive background fields
we have to consider the components $g_{ab}$ as external fields.
Such a treatment corresponds to covariant methods where
functional integral is calculated from the very beginning only
over $X^\mu$ variables.

After the standard parametrization of metric
\begin{eqnarray}
g_{ab}& = &e^{\gamma}\left(\begin{array}{cc}
\lambda^{2}_{1}-\lambda^{2}_{0} & \lambda_{1} \\
\lambda_{1}  & 1 \end{array}\right)
\end{eqnarray}
the hamiltonian in linear approximation in background fields
takes the form

\begin{equation}
H=\int d\sigma\,(p_\mu {\dot x}^\mu - L)
=\int d\sigma \,(\lambda_{0}T_{0}+\lambda_{1}T_{1}),
\label{H}
\end{equation}
where
\begin{eqnarray}
& & T_0 = T_0^{(0)} + T_0^{(1)},
\nonumber \\
& & T_0^{(0)} = {1\over2}\left(2\pi\alpha'P^2 +
(1/2\pi\alpha) X'^2 \right),
\nonumber \\
& & T_0^{(1)} = {1\over 2\pi} \biggl( (1/\alpha') e^{\gamma}Q(X)
+  (1/\alpha') e^{-\gamma}
\Bigl[ (2\pi\alpha')^4 P^\mu P^\nu P^\lambda P^\kappa
\nonumber \\& &
\qquad\qquad\qquad\qquad
{}- 2 (2\pi\alpha')^2 P^\mu P^\nu X'^\lambda X'^\kappa
+ X'^\mu X'^\nu X'^\lambda X'^\kappa \Bigr] F^1_{\mu\nu\lambda\kappa}(X)
\nonumber \\& &
\qquad\qquad
{}+ (2/\alpha')e^{-\gamma}
\Bigl[ - (2\pi\alpha')^3 P^\mu P^\nu P^\lambda X'^\kappa
+  2\pi\alpha' X'^\mu X'^\nu P^\lambda X'^\kappa
    \Bigr]  F^2_{\mu\nu\lambda\kappa}(X)
\nonumber \\& &
\qquad\qquad
{}+ R^{(2)} \Bigl[ - (2\pi\alpha')^2 P^\mu P^\nu +
X'^\mu X'^\nu \Bigr] W^1_{\mu\nu}(X)
+ 4\pi\alpha' R^{(2)}  P^\mu X'^\nu W^2_{\mu\nu}(X)
\nonumber \\& &
\qquad\qquad
{} + \alpha' e^\gamma R^{(2)} R^{(2)} C(X)
  \biggr) ,
\nonumber \\
& & T_1 = T_1^{(0)} = P_{\mu}X'^\mu {.}
\label{T}
\end{eqnarray}
$X'^\mu=\partial X^\mu/\partial\sigma$, and $P_\mu$ are momenta canonically
conjugated to $X^\mu$.
$T_0^{(0)}$ and $T_1^{(0)}$ represent constraints of free string
theory and form closed algebra in terms of Poisson brackets.
$\lambda_0$ and $\lambda_1$ play the role of external fields and
so $T_0$ and  $T_1$ cannot be considered as constraints of
classical theory. In free string theory conditions
$T^{(0)}_0=0$, $T^{(0)}_1=0$ result from conservation of
canonical momenta conjugated to $\lambda_0$ and $\lambda_1$.
According to our prescription in string theory with massive
background fields  $\lambda_0$ and $\lambda_1$ can not be
considered as dynamical variables, there are no corresponding
momenta and conditions of their conservation do not appear.

The role of parameters $a$ in the theory under consideration is
played by background fields $Q$, $F$, $W$, $C$ and conformal
factor of two-dimensional metric $\gamma(\tau,\sigma)$. The
theory (\ref{H}) is of the type (\ref{Hgen}) with $H_0=0$,
structural constants of classical algebra being independent on
time. It means that the condition of conservation for the
operator $\Omega$ (\ref{Omega}) will be satisfied if $T_0$
does not depend on time explicitly, that is
\begin{equation}
{\partial\gamma(\tau,\sigma)\over\partial\tau}=0, \qquad
{\partial R^{(2)}(\tau,\sigma)\over\partial\tau}=0 {.}
\label{gamma}
\end{equation}

To derive the first condition (\ref{a_0}) one has to construct
quantum algebra of operators $T_0$, $T_1$.

\section{Quantum guage algebra and effective equations
of motion}

To carry out the calculation of the quantum algebra we first of
all introduce the following notations for so called right and
left moving Fubini fields
\begin{equation}
Y^{\pm\mu} = 2\pi\alpha' P^\mu \mp X'^\mu
\end{equation}
and turn to the discrete set of operators:
\begin{eqnarray}
L_n & = & \int^{2\pi}_0 d\sigma\, e^{-in\sigma}
{1\over 2} (T_0-T_1) = L_n^{(0)} + K_n {,}
\nonumber \\
\bar{L}_n & = & \int^{2\pi}_0 d\sigma\, e^{in\sigma}
{1\over 2} (T_0+T_1) = \bar{L}_n^{(0)} - K_{-n} {.}
\end{eqnarray}
Here operators
\begin{eqnarray}
& &L_n^{(0)}=\int^{2\pi}_0 d\sigma\, e^{-in\sigma}
{1\over 2}(T_0^{(0)}-T_1^{(0)})
=\int^{2\pi}_0 d\sigma\, e^{-in\sigma}
{1\over 8\pi\alpha'} Y^{+\mu} Y^{+\nu} \eta_{\mu\nu},
\nonumber \\
& &\bar{L}_n^{(0)}=\int^{2\pi}_0 d\sigma\, e^{in\sigma}
{1\over 2}(T_0^{(0)}+T_1^{(0)})
=\int^{2\pi}_0 d\sigma\, e^{in\sigma}
{1\over 8\pi\alpha'} Y^{-\mu} Y^{-\nu} \eta_{\mu\nu}
\end{eqnarray}
represent the set of constraints of the free string theory and
form the well known quantum algebra
\begin{eqnarray}
[L_n^{(0)} , L_m^{(0)} ] & = & \hbar (n-m)L_{n+m}^{(0)} +
\hbar^2 \delta_{n,-m}  \left( {D\over 12}n(n^2 -1) + 2\alpha(0)n \right),
\nonumber \\
{} [\bar{L}_n^{(0)} , \bar{L}_m^{(0)} ] & = & \hbar (n-m)\bar{L}_{n+m}^{(0)} +
\hbar^2 \delta_{n,-m}  \left( {D\over 12}n(n^2 -1) + 2\beta(0)n \right),
\nonumber \\
{} [L_n^{(0)} , \bar{L}_m^{(0)} ] & = & 0,
\end{eqnarray}
$\alpha(0)$, $\beta(0)$ are constants fixing the operators
ordering ambiguity.

The operators $K_n$ are contributions linear in background
fields and have the form
\begin{eqnarray}
& & K_n = {1\over 4\pi\alpha'} \int^{2\pi}_0 d\sigma\,
e^{-in\sigma} \Bigl\{ e^\gamma Q(X) +
e^{-\gamma} Y^{+\mu}Y^{+\nu}Y^{-\lambda}Y^{-\kappa}
			   F_{\mu\nu\lambda\kappa}(X)
\nonumber \\
& & \qquad\qquad {} + \alpha' R^{(2)} Y^{+\mu}Y^{-\nu} W_{\mu\nu}(X)
+ (\alpha')^2 e^\gamma R^{(2)} R^{(2)} C(X) \Bigr\}
\end{eqnarray}
where we introduced the following notations:
\begin{eqnarray}
& & F_{\mu\nu\lambda\kappa} =
2 F^1_{\mu\lambda\nu\kappa} + 2 F^1_{\mu\kappa\nu\lambda} -
2 F^2_{\mu\lambda\nu\kappa} - 2 F^2_{\nu\kappa\mu\lambda} {,}
\nonumber \\
& &W_{\mu\nu}=-W^1_{\mu\nu}+W^2_{\mu\nu}{.}
\end{eqnarray}

As periodic functions of $\sigma$ the canonical operators of
string coordinates and momenta have standard Fourier expansions:
\begin{eqnarray}
X^{\mu} & = & \frac{1}{\sqrt{2\pi}}x_0^{\mu}(\tau) +
i\sqrt{\frac{\alpha'}{2}} \sum_{n\neq 0} \frac{1}{n}
(\alpha^{\mu}_{n} (\tau) z^n
+ \bar{\alpha}{}^{\mu}_n (\tau) z^{-n}), \quad z=e^{i\sigma},
\nonumber\\
P^{\mu} & = & \frac{1}{\sqrt{2\pi\alpha'}} p_0^\mu (\tau)
+ \frac{1}{2\pi\sqrt{2\alpha'}} \sum_{n\neq 0}
(\alpha^{\mu}_{n} (\tau) z^n
+ \bar{\alpha}{}^{\mu}_n (\tau) z^{-n})
\end{eqnarray}
where operators of zero modes $x_0^\mu$, $p_0^\mu$ and
oscillating ones $\alpha^\mu_n$, $\bar{\alpha}{}^\mu_n$
satisfy the following commutation relations (we have set
$\hbar=1$ for the rest of the paper):
\begin{equation}
[x_0^\mu, p_{0\nu}]=i \delta_\nu^\mu, \;
[\alpha^\mu_m, \alpha^\nu_n]=
[\bar{\alpha}{}^\mu_m,\bar{\alpha}{}^\nu_n]=
m \delta_{m,-n}\eta^{\mu\nu} {.}
\end{equation}

The most general ordering in terms of the operators
$x_0^\mu$, $p_0^\mu$, $\alpha^\mu_n$, $\bar{\alpha}{}^\mu_n$ can
be written down as
\begin{eqnarray}
& & O(\alpha^\mu_m \alpha^\nu_n) =
(1-c_m^\mu)\alpha^\mu_m \alpha^\nu_n
+ c_m^\mu \alpha^\nu_n \alpha^\mu_m =
\alpha^\mu_m \alpha^\nu_n  - m c_m^\mu
\delta_{m,-n}\eta^{\mu\nu} {,}
\nonumber\\
& & O(\bar{\alpha}{}^\mu_m \bar{\alpha}{}^\nu_n) =
(1-\bar{c}{}_m^\mu)
\bar{\alpha}{}^\mu_m \bar{\alpha}{}^\nu_n +
\bar{c}{}_m^\mu \bar{\alpha}{}^\nu_n \bar{\alpha}{}^\mu_m =
\bar{\alpha}{}^\mu_m \bar{\alpha}{}^\nu_n
- m \bar{c}_m^\mu \delta_{m,-n}\eta^{\mu\nu} {,}
\nonumber \\
& & O(x^\mu_0 p_{\nu 0}) =
(1-c_0^\mu) x^\mu_0 p_{\nu 0} + c_0^\mu p_{\nu 0} x^\mu_0 =
x^\mu_0 p_{\nu 0} - i c_0^\mu \delta^\mu_\nu {.}
\end{eqnarray}
Here $c_m^\mu$, $\bar{c}{}_m^\mu$, $c_0^\mu$ are some constant
parameters defining an ordering type and there are no summations
over indices $m$ and $\mu$.

As one can commute operators within the sign of ordering
\begin{equation}
O(\alpha^\mu_m \alpha^\mu_{-m}) = O(\alpha^\mu_{-m} \alpha^\mu_m)
\end{equation}
then there exist the following relation for parameters
$c_m^\mu$:
\begin{equation}
c_{-m}^\mu + c_m^\mu = 1
\end{equation}
and the analogous one for $\bar{c}{}_m^\mu$. It means that
operators ordering is fixed up completely by choosing values for
the infinite set of parameters $c_m^\mu$, $\bar{c}{}_m^\mu$,
$c_0^\mu$, $m>0$. For example, the special case of Wick ordering
(that is such an ordering when all the creation operators
$\alpha_n$, $\bar{\alpha}{}_n$, $n<0$ stand to the right of all the
annihilating operators $\alpha_n$, $\bar{\alpha}{}_n$, $n>0$) is
defined by the choosing $c_n^\mu=\theta(n)$:
\begin{equation}
:\alpha^\mu_m \alpha^\nu_n: =
\alpha^\mu_m \alpha^\nu_n  - m \theta(m) \delta_{m,-n}\eta^{\mu\nu}{.}
\end{equation}

Contractions of fundamental operators have the form:
$$
\stackrel
{\displaystyle
\alpha_m^\mu \alpha_n^\nu =
\alpha_m^\mu \alpha_n^\nu - O(\alpha_m^\mu \alpha_n^\nu) =
m c_m^\mu \delta_{m,-n} \eta^{\mu\nu} {,}}
{\rule{0.4pt}{1.5mm}
\rule{6mm}{1pt}
\rule{0.4pt}{1.5mm}
\phantom{{}_n^\nu =
\alpha_m^\mu \alpha_n^\nu - O(\alpha_m^\mu \alpha_n^\nu) =
m c_m^\mu \delta_{m,-n} \eta^{\mu\nu} {,}}}
$$
\begin{equation}
\stackrel
{\displaystyle
\bar{\alpha}{}_m^\mu \bar{\alpha}{}_n^\nu =
m \bar{c}{}_m^\mu \delta_{m,-n} \eta^{\mu\nu} {,}}
{\rule{0.4pt}{1.5mm}
\rule{6mm}{1pt}
\rule{0.4pt}{1.5mm}
\phantom{
{}_n^\nu = m \bar{c}{}_m^\mu \delta_{m,-n} \eta^{\mu\nu} {,}}}
\quad
\stackrel
{\displaystyle
x_0^\mu p_{0 \nu} = i c_0^\mu \delta^\mu_\nu {,}}
{\rule{0.4pt}{1.5mm}
\rule{4mm}{1pt}
\rule{0.4pt}{1.5mm}
\phantom{{}_{0 \nu} = i c_0^\mu \delta^\mu_\nu {,}}}
\quad
\stackrel
{\displaystyle
p_{0 \nu} x_0^\mu = i (c_0^\mu-1) \delta^\mu_\nu {.}}
{\rule{0.4pt}{1.5mm}
\rule{6mm}{1pt}
\rule{0.4pt}{1.5mm}
\phantom{{}_0^\mu = i (c_0^\mu-1) \delta^\mu_\nu {.}}}
\end{equation}

In the case of free string theory the whole ordering ambiguity of
constraints operators reduces to the choice of two parameters in the
following way. Operators $L_n^{(0)}$, $\bar{L}{}_n^{(0)}$ are
quadratic in fundamental variables
\begin{equation}
L_n^{(0)} = \frac{1}{2} \sum_k
\alpha_k^\mu \alpha_{n-k}^\nu \eta_{\mu\nu} {,} \quad
\bar{L}{}_n^{(0)} = \frac{1}{2} \sum_k
\bar{\alpha}{}_k^\mu \bar{\alpha}{}_{n-k}^\nu \eta_{\mu\nu}
\end{equation}
and their arbitrary ordering can be expressed through the Wick
one as follows:
\begin{eqnarray}
O(L_n^{(0)}) &=& \frac{1}{2} \sum_{k=-\infty}^\infty
:\alpha_k^\mu \alpha_{n-k}^\nu \eta_{\mu\nu}:
- \frac{1}{2} \sum_{k=-\infty}^\infty
 \sum_{\mu=0}^{D-1} k\delta_{n,0} c_k^\mu
+ \frac{1}{2} \sum_{k=-\infty}^\infty
 \sum_{\mu=0}^{D-1} k\delta_{n,0} \theta(k)
\nonumber\\
&=&:L_n^{(0)}: - \alpha(0) \delta_{n,0} {,}
\nonumber\\
O(\bar{L}{}_n^{(0)}) &=& :\bar{L}{}_n^{(0)}:
 - \beta(0) \delta_{n,0}  {,}
\end{eqnarray}
where parameters $\alpha(0)$ and $\beta(0)$ are defined as
\begin{equation}
\alpha(0) = \sum_{k=1}^\infty \sum_{\mu=0}^{D-1} k (c_k^\mu-1)
 {,}\quad
\beta(0) = \sum_{k=1}^\infty \sum_{\mu=0}^{D-1} k
(\bar{c}{}_k^\mu-1) {.}
\label{alpha(0)}
\end{equation}
Any set of parameters $c_n^\mu$, $\bar{c}_n^\mu$ leading to the same values
of $\alpha(0)$ and $\beta(0)$ defines equivalent quantum free string
theories. Parameters $c_0^\mu$ do not enter the operators
$L_n^{(0)}$, $\bar{L}{}_n^{(0)}$ at all and so $\alpha(0)$ and
$\beta(0)$ are the only ordering parameters whose values are
relevant in free string theory.

In the case of string interacting with background fields the
operators $L_n$, $\bar{L}{}_n$ are arbitrary functions of the
operators $x_0^\mu$, $p_0^\mu$, $\alpha^\mu_n$,
$\bar{\alpha}{}^\mu_n$  and one has to take into account the whole
infinite set of various parameters $c_n^\mu$, $\bar{c}_n^\mu$,
$c_0^\mu$. We will
calculate the quantum algebra of operators $L_n$, $\bar{L}{}_n$ using
ordering parameters of the following form
\begin{eqnarray}
& &c_k^\mu-1 = \frac{\alpha(0)}{D} (1-\Lambda )^2 \Lambda ^{k-1},
\quad k>0, \quad |\Lambda |<1,
\nonumber \\
& & \bar{c}{}_k^\mu-1 = \frac{\beta(0)}{D} (1-\bar \Lambda )^2
\bar{\Lambda }^{k-1}, \quad k>0, \quad |\bar \Lambda |<1,
\nonumber \\
& & c_0^\mu = \frac{1}{2} {,}
\end{eqnarray}
where $\Lambda $, $\bar \Lambda $ are complex variables. This choice is not the
most general one but it is suitable for our purposes because it both
is consistent with the free string ordering prescription
(\ref{alpha(0)}) and by
means of parameters $\Lambda $, $\bar \Lambda $ describes an ordering ambiguity
which was not relevant in the free case but is sufficient for the
string in background fields.

In this class of ordering prescription the contractions of
fundamental variables have the form
\begin{eqnarray}
& &
\stackrel
{\displaystyle
X^\mu(z_1)Y^{+\nu}(z_2) = {}}
{\rule{0.4pt}{1.5mm}
\rule{12mm}{1pt}
\rule{0.4pt}{1.5mm}
\phantom{{}^{+\nu}(z_2) = {}}}
\nonumber\\
& & \quad
{}=i\alpha'\eta^{\mu\nu} \left[ \frac{1}{2}+\sum_{n>0}
\left(\frac{z_1}{z_2}\right)^n \right]
+ i\alpha'\eta^{\mu\nu}\frac{\alpha(0)(1-\Lambda )^2}{\Lambda D}  \sum_{n>0}
\left[ \left(\frac{\Lambda z_1}{z_2}\right)^n
+\left(\frac{\Lambda z_2}{z_1}\right)^n  \right] {,}
\nonumber \\
& &
\stackrel
{\displaystyle
X^\mu(z_1)Y^{-\nu}(z_2) = {}}
{\rule{0.4pt}{1.5mm}
\rule{12mm}{1pt}
\rule{0.4pt}{1.5mm}
\phantom{{}^{-\nu}(z_2) = {}}}
\nonumber\\
& & \quad
{}=-i\alpha'\eta^{\mu\nu} \left[ \frac{1}{2}+\sum_{n>0}
\left(\frac{z_2}{z_1}\right)^n \right]
- i\alpha'\eta^{\mu\nu}\frac{\beta(0)(1-\bar{\Lambda })^2}{\bar{\Lambda }D}  \sum_{n>0}
\left[ \left(\frac{\bar{\Lambda }z_1}{z_2}\right)^n
+\left(\frac{\bar{\Lambda }z_2}{z_1}\right)^n  \right] {,}
\nonumber\\
& &
\stackrel
{\displaystyle
Y^{+\mu}(z_1)Y^{+\nu}(z_2) = {}}
{\rule{0.4pt}{1.5mm}
\rule{13mm}{1pt}
\rule{0.4pt}{1.5mm}
\phantom{{}^{+\nu}(z_2) = {}}}
\nonumber\\
& & \quad
{}=\alpha'\eta^{\mu\nu} \sum_{n>0} n \left(\frac{z_1}{z_2}\right)^n
+ 2\alpha'\eta^{\mu\nu}\frac{\alpha(0)(1-\Lambda )^2}{\Lambda D}  \sum_{n>0} n
\left[ \left(\frac{\Lambda z_1}{z_2}\right)^n
+\left(\frac{\Lambda z_2}{z_1}\right)^n  \right] {,}
\nonumber\\
& &\stackrel
{\displaystyle
Y^{-\mu}(z_1)Y^{-\nu}(z_2) = {}}
{\rule{0.4pt}{1.5mm}
\rule{13mm}{1pt}
\rule{0.4pt}{1.5mm}
\phantom{{}^{-\nu}(z_2) = {}}}
\nonumber\\
& & \quad
{}=\alpha'\eta^{\mu\nu} \sum_{n>0} n \left(\frac{z_2}{z_1}\right)^n
+ 2\alpha'\eta^{\mu\nu}\frac{\beta(0)(1-\bar{\Lambda })^2}{\bar{\Lambda }D}  \sum_{n>0} n
\left[ \left(\frac{\bar{\Lambda }z_1}{z_2}\right)^n
+\left(\frac{\bar{\Lambda }z_2}{z_1}\right)^n  \right] {,}
\nonumber\\
& &\stackrel
{\displaystyle
Y^{+\mu}(z_1)Y^{-\nu}(z_2) = 0}
{\rule{0.4pt}{1.5mm}
\rule{13mm}{1pt}
\rule{0.4pt}{1.5mm}
\phantom{{}^{-\nu}(z_2) = 0}} {.}
\end{eqnarray}

All the contractions contain divergent series and represent
generalized functions of their argument $(z_1-z_2)$. Quantum
contributions to the operator algebra contain products of these
contractions and so one should regularize them by introducing a
parameter which would make the contractions well defined
analytical functions.

We introduce this parameter $\epsilon$ ($Re \, \epsilon>0$) in such a way
that shifts arguments of divergent series into the region of
their convergence:
$$
\frac{z_1}{z_2} \longrightarrow e^{-\epsilon} \frac{z_1}{z_2}{,} \quad
\frac{z_2}{z_1} \longrightarrow e^{-\epsilon} \frac{z_2}{z_1}{.}
$$
Then all the contractions can be summed up to elementary analytical
functions:
\begin{eqnarray}
& &
\stackrel
{\displaystyle
X^\mu(z_1)Y^{+\nu}(z_2) = {}}
{\rule{0.4pt}{1.5mm}
\rule{12mm}{1pt}
\rule{0.4pt}{1.5mm}
\phantom{{}^{+\nu}(z_2) = {}}}
\nonumber\\
& & \quad
{}=i\alpha'\eta^{\mu\nu} \left[ \frac{1}{2}+
\frac{z_1 e^{-\epsilon}}{z_2-z_1 e^{-\epsilon}} \right]
+ i\alpha'\eta^{\mu\nu}\frac{\alpha(0)(1-\Lambda )^2}{\Lambda D}
\left[ \frac{z_1 \Lambda }{z_2-z_1 \Lambda }+\frac{z_2}{z_2-z_1 \Lambda ^{-1}}  \right] {,}
\nonumber \\
& &
\stackrel
{\displaystyle
X^\mu(z_1)Y^{-\nu}(z_2) = {}}
{\rule{0.4pt}{1.5mm}
\rule{12mm}{1pt}
\rule{0.4pt}{1.5mm}
\phantom{{}^{-\nu}(z_2) = {}}}
\nonumber\\
& & \quad
{}=-i\alpha'\eta^{\mu\nu} \left[ \frac{1}{2}+
\frac{z_1 e^{\epsilon}}{z_2-z_1 e^{\epsilon}} \right]
- i\alpha'\eta^{\mu\nu}\frac{\beta(0)(1-\bar{\Lambda })^2}{\bar{\Lambda }D}
\left[ \frac{z_1 \bar{\Lambda }}{z_2-z_1 \bar{\Lambda }}
+\frac{z_2}{z_2-z_1 \bar{\Lambda }^{-1}}  \right] {,}
\nonumber \\
& &
\stackrel
{\displaystyle
Y^{+\mu}(z_1)Y^{+\nu}(z_2) = {}}
{\rule{0.4pt}{1.5mm}
\rule{13mm}{1pt}
\rule{0.4pt}{1.5mm}
\phantom{{}^{+\nu}(z_2) = {}}}
\nonumber\\
& & \quad
{}=\alpha'\eta^{\mu\nu}
\frac{z_1 z_2 e^{-\epsilon}}{(z_2-z_1 e^{-\epsilon})^2}
+ 2\alpha'\eta^{\mu\nu}\frac{\alpha(0)(1-\Lambda )^2}{\Lambda D}
\left[ \frac{z_1 z_2 \Lambda }{(z_2-z_1 \Lambda )^2}+
\frac{z_2 z_1 \Lambda ^{-1}}{(z_2-z_1 \Lambda ^{-1})^2}  \right] {,}
\nonumber \\
& &
\stackrel
{\displaystyle
Y^{-\mu}(z_1)Y^{-\nu}(z_2) = {}}
{\rule{0.4pt}{1.5mm}
\rule{13mm}{1pt}
\rule{0.4pt}{1.5mm}
\phantom{{}^{-\nu}(z_2) = {}}}
\nonumber\\
& & \quad
{}=\alpha'\eta^{\mu\nu}
\frac{z_1 z_2 e^{\epsilon}}{(z_2-z_1 e^{\epsilon})^2}
+ 2\alpha'\eta^{\mu\nu}\frac{\beta(0)(1-\bar{\Lambda })^2}{\bar{\Lambda }D}
\left[ \frac{z_1 z_2 \bar{\Lambda }}{(z_2-z_1 \bar{\Lambda })^2}+
\frac{z_2 z_1 \bar{\Lambda }^{-1}}{(z_2-z_1 \bar{\Lambda }^{-1})^2}  \right]
\nonumber\\
\end{eqnarray}

Integrations in operators $L_n$, $\bar{L}{}_n$ go over the unit
circle in complex plane of the variable $z=e^{i\sigma}$. Our
regularization scheme leads to disappearing of the singular point
$z_2=z_1$ on the contour of integration and to appearing of
additional singular points $z_2=z_1 e^{-\epsilon}$, $z_2=z_1
e^\epsilon$ lying out of the contour. Therefore, all the
integrals over $z_2$ can be done by means of calculating
residues in internal poles $z_2=0$,  $z_2=z_1 e^{-\epsilon}$,
$z_2=z_1 \Lambda $, $z_2=z_1 \bar\Lambda $.
The regularization parameter $\epsilon$ should be
set equal to zero after doing all calculations.

Now we are ready to calculate commutators $[L_n^{(0)}+K_n,
L_m^{(0)}+K_m]$. We are interested only in contributions linear
in background fields and so can omit the commutator $[K_n, K_m]$
because each term in operators $K_n$ depends on background
fields.  Operators $L_n^{(0)}$ are quadratic and so
commutators $[K_n, L_m^{(0)}]$ contain quantum contributions
only with products of two contractions.

The general expression for these commutators can be constructed
with the use of Wick theorem for product of two operators
$A(\Gamma)$ and $B(\Gamma)$ \cite{berezin}:
\begin{equation}
\stackrel
{\displaystyle
O(A(\Gamma)) \, O(B(\Gamma)) =
O \Bigl( \exp \Bigl(
\Gamma_1^M \Gamma_2^N
\frac{\delta~}{\delta \Gamma_1^M} \frac{\delta~}{\delta \Gamma_2^N} \Bigr)
A(\Gamma_1) B(\Gamma_2) \Bigr) \Bigr|_{\Gamma_1=\Gamma_2=\Gamma}}
{\hspace{5cm}\rule{0.4pt}{1.5mm}
\rule{6mm}{1pt}
\rule{0.4pt}{1.5mm}\hspace{6cm}}
\end{equation}
Set of canonical variables $\Gamma$ in our case consists
of the operators $X^\mu(\tau,z)$, $Y^{+\mu}(\tau,z)$,
$Y^{-\mu}(\tau,z)$. After decomposition of canonical
contractions into the sum of symmetrical and antisymmetrical
parts
\begin{equation}
\stackrel
{\displaystyle
\Gamma^M \Gamma^N = \Gamma^M \Gamma^N + \Gamma^M \Gamma^N}
{\rule{0.4pt}{1.5mm}\rule{6mm}{1pt}\rule{0.4pt}{1.5mm}
\phantom{{}^N = \Gamma}
\rule{0.4pt}{1.5mm}\rule{6mm}{1pt}\rule{0.4pt}{1.5mm}~{}_S
\phantom{{} + \Gamma}
\rule{0.4pt}{1.5mm}\rule{6mm}{1pt}\rule{0.4pt}{1.5mm}~{}_A}
\end{equation}
commutator of two operators according to the Wick theorem takes
form of the following expansion in powers of contractions:
\begin{equation}
\stackrel
{\displaystyle
[A,B] = 2\Gamma^M \Gamma^N
\frac{\delta A}{\delta \Gamma^M} \frac{\delta B}{\delta \Gamma^N}
+ 4\Gamma^{M_1} \Gamma^{N_1} \Gamma^{M_2} \Gamma^{N_2}
\frac{\delta^2 A}{\delta\Gamma^{M_1}\delta\Gamma^{M_2}}
\frac{\delta^2 B}{\delta\Gamma^{N_1}\delta\Gamma^{N_2}} + \ldots}
{\hspace{18mm}
\rule{0.4pt}{1.5mm}\rule{6mm}{1pt}\rule{0.4pt}{1.5mm}~_A
\hspace{26mm}
\rule{0.4pt}{1.5mm}\rule{6mm}{1pt}\rule{0.4pt}{1.5mm}~_S~~~
\rule{0.4pt}{1.5mm}\rule{6mm}{1pt}\rule{0.4pt}{1.5mm}~_A
\hspace{5cm}}
\end{equation}
The first term reproduces classical contributions proportional to
Poisson bracket of the corresponding classical variables.

In our case the operator $A=\oint(dz/iz)a(z)$ depends on
all the canonical variables $X^\mu$, $Y^{+\mu}$, $Y^{-\mu}$ and
the operator $B=\oint(dz/iz)b(z)$ depends only on
$Y^{+\mu}$ and so the contribution quadratic in contractions
$\Delta^{(2)}$ in the commutator is
\begin{eqnarray}
& &
\stackrel
{\displaystyle
\Delta^{(2)} = \oint\frac{dz_1}{iz_1} \oint\frac{dz_2}{iz_2}
\biggl\{2 X^{\mu_1}(z_1) Y^{+\nu_1}(z_2) X^{\mu_2}(z_1) Y^{+\nu_2}(z_2)}
{\hspace{40mm}
\rule{0.4pt}{1.5mm}\rule{13mm}{1pt}\rule{0.4pt}{1.5mm}~_A
\hspace{12mm}
\rule{0.4pt}{1.5mm}\rule{15mm}{1pt}\rule{0.4pt}{1.5mm}~_S
\hspace{7mm}}
\nonumber\\
& &\hspace{5cm}{}\times
\frac{\partial^2 a}{\partial X^{\mu_1} \partial X^{\mu_2}} (z_1)
\frac{\partial^2 b}{\partial Y^{+\nu_1}  \partial  Y^{+\nu_2}} (z_2)
\nonumber\\
& &
\stackrel
{\displaystyle
{}+2 \Bigl(X^{\mu_1}(z_1) Y^{+\nu_1}(z_2) Y^{+\mu_2}(z_1) Y^{+\nu_2}(z_2)
+ X^{\mu_1}(z_1) Y^{+\nu_1}(z_2) Y^{+\mu_2}(z_1) Y^{+\nu_2}(z_2)\Bigr)}
{\hspace{10mm}
\rule{0.4pt}{1.5mm}\rule{13mm}{1pt}\rule{0.4pt}{1.5mm}~_A
\hspace{12mm}
\rule{0.4pt}{1.5mm}\rule{15mm}{1pt}\rule{0.4pt}{1.5mm}~_S
\hspace{16mm}
\rule{0.4pt}{1.5mm}\rule{13mm}{1pt}\rule{0.4pt}{1.5mm}~_S
\hspace{12mm}
\rule{0.4pt}{1.5mm}\rule{15mm}{1pt}\rule{0.4pt}{1.5mm}~_A
\hspace{1cm}}
\nonumber\\
& &\hspace{5cm}{}\times
\frac{\partial^2 a}{\partial X^{\mu_1} \partial Y^{+\mu_2}} (z_1)
\frac{\partial^2 b}{\partial Y^{+\nu_1}  \partial Y^{+\nu_2}} (z_2)
\nonumber\\
& &
\stackrel
{\displaystyle
{}+Y^{+\mu_1}(z_1) Y^{+\nu_1}(z_2) Y^{+\mu_2}(z_1) Y^{+\nu_2}(z_2)
\frac{\partial^2 a}{\partial Y^{+\mu_1} \partial Y^{+\mu_2}} (z_1)
\frac{\partial^2 b}{\partial Y^{+\nu_1}  \partial Y^{+\nu_2}} (z_2)
\biggr\} {.} }
{\hspace{6mm}
\rule{0.4pt}{1.5mm}\rule{15mm}{1pt}\rule{0.4pt}{1.5mm}~_A
\hspace{12mm}
\rule{0.4pt}{1.5mm}\rule{15mm}{1pt}\rule{0.4pt}{1.5mm}~_S
\hspace{75mm}}
\end{eqnarray}

Substituting the operators $K_n$, $L_n^{(0)}$ for $A$, $B$ in
this formula and doing all integrations over $z_2$ with the use
of regularized contractions we arrive at the following
commutators:
\begin{eqnarray}
& & [L_n , L_m ] = (n-m)L_{n+m} +
\delta_{n,-m}  \left( {D\over 12}n(n^2 -1) + 2\alpha(0)n \right)
\nonumber \\ {}& & \qquad
{}+ \int^{2\pi}_0 d\sigma\, e^{-i(n+m)\sigma} \Biggl\{
{(m-n)\over 16\pi} \biggl[ e^\gamma (\partial^2 + 4/\alpha') Q(X)
\nonumber \\ {}& & \qquad
{} +  e^{-\gamma} Y^{+\mu} Y^{+\nu}Y^{-\lambda}Y^{-\kappa}
(\partial^2-4/\alpha') F_{\mu\nu\lambda\kappa}(X)
\nonumber \\ {}& & \qquad
{}+ \alpha' R^{(2)} Y^{+\mu} Y^{-\nu} \partial^2 W_{\mu\nu}(X)
+ \alpha'^2 e^\gamma R^{(2)} R^{(2)} (\partial^2 + 4/\alpha') C(X) \biggr]
\nonumber \\ {}& & \qquad
{}+ {i(m^2-n^2)\over 8\pi} \biggl[
2 e^{-\gamma} Y^{+\nu}Y^{-\lambda}Y^{-\kappa}
\partial^\mu F_{\mu\nu\lambda\kappa}(X)
+ \alpha' R^{(2)} Y^{-\nu} \partial^\mu W_{\mu\nu}(X) \biggr]
\nonumber \\ {}& & \qquad
{}+ \biggl[{m-n+n^3-m^3\over 24\pi}+{\alpha(0)(n-m)\over\pi D} \biggr]
Y^{-\lambda}Y^{-\kappa}  F^\mu{}_{\mu\lambda\kappa}(X) \Biggr\} {,}
\end{eqnarray}
Analogous calculations give the other commutators:
\begin{eqnarray}
& & [\bar{L}_n , \bar{L}_m ] =  (n-m)\bar{L}_{n+m} +
 \delta_{n,-m}  \left( {D\over 12}n(n^2 -1) + 2\beta(0)n \right)
\nonumber \\ {}& & \qquad
{}+ \int^{2\pi}_0 d\sigma\, e^{i(n+m)\sigma} \Biggl\{
{(m-n)\over 16\pi} \biggl[ e^\gamma (\partial^2 + 4/\alpha') Q(X)
\nonumber \\ {}& & \qquad
{} +  e^{-\gamma} Y^{+\mu} Y^{+\nu}Y^{-\lambda}Y^{-\kappa}
(\partial^2-4/\alpha') F_{\mu\nu\lambda\kappa}(X)
\nonumber \\ {}& & \qquad
{}+ \alpha' R^{(2)} Y^{+\mu} Y^{-\nu} \partial^2 W_{\mu\nu}(X)
+ \alpha'^2 e^\gamma R^{(2)} R^{(2)} (\partial^2 + 4/\alpha') C(X) \biggr]
\nonumber \\ {}& & \qquad
{}+ {i(m^2-n^2)\over 8\pi} \biggl[
2 e^{-\gamma} Y^{+\mu}Y^{+\nu}Y^{-\kappa}
\partial^\lambda F_{\mu\nu\lambda\kappa}(X)
+ \alpha' R^{(2)} Y^{+\mu} \partial^\nu W_{\mu\nu}(X) \biggr]
\nonumber \\ {}& & \qquad
{}+ \biggl[{m-n+n^3-m^3\over 24\pi}+{\beta(0)(n-m)\over\pi D} \biggr]
Y^{+\mu}Y^{+\nu}  F_{\mu\nu}{}^\lambda{}_\lambda(X) \Biggr\} {,}
\nonumber \\
& & [L_n , \bar{L}_m ] =
 \int^{2\pi}_0 d\sigma\, e^{i(m-n)\sigma} \Biggl\{
{(m-n)\over 16\pi} \biggl[ e^\gamma (\partial^2 + 4/\alpha') Q(X)
\nonumber \\ {}& & \qquad
{} +  e^{-\gamma} Y^{+\mu} Y^{+\nu}Y^{-\lambda}Y^{-\kappa}
(\partial^2-4/\alpha') F_{\mu\nu\lambda\kappa}(X)
\nonumber \\ {}& & \qquad
{}+ \alpha' R^{(2)} Y^{+\mu} Y^{-\nu} \partial^2 W_{\mu\nu}(X)
+ \alpha'^2 e^\gamma R^{(2)} R^{(2)} (\partial^2 + 4/\alpha') C(X) \biggr]
\nonumber \\ {}& & \qquad
{}- {n^2+4\alpha(0)/D \over 8\pi} \biggl[
2 e^{-\gamma} Y^{+\nu}Y^{-\lambda}Y^{-\kappa}
\partial^\mu F_{\mu\nu\lambda\kappa}(X)
+ \alpha' R^{(2)} Y^{-\nu} \partial^\mu W_{\mu\nu}(X) \biggr]
\nonumber \\ {}& & \qquad
{}+ {m^2+4\beta(0)/D \over 8\pi} \biggl[
2 e^{-\gamma} Y^{+\mu}Y^{+\nu}Y^{-\kappa}
\partial^\lambda F_{\mu\nu\lambda\kappa}(X)
+ \alpha' R^{(2)} Y^{+\mu} \partial^\nu W_{\mu\nu}(X) \biggr]
\nonumber \\ {}& & \qquad
{}-\biggl({n-n^3\over 6}-{4n\alpha(0)\over D}+{2\alpha(0)\over D}\biggr)
{1\over 4\pi}
Y^{-\lambda}Y^{-\kappa}  F^\mu{}_{\mu\lambda\kappa}(X)
\nonumber \\ {}& & \qquad
{}+\biggl({m-m^3\over 6}-{4m\beta(0)\over D}+{2\beta(0)\over D}\biggr)
{1\over 4\pi}
Y^{+\mu}Y^{+\nu}  F_{\mu\nu}{}^\lambda{}_\lambda(X)
\nonumber \\ {}& & \qquad
{}-{i\over 4\pi\alpha'} \biggl[ (e^\gamma)' Q(X)
+  (e^{-\gamma})' Y^{+\mu} Y^{+\nu}Y^{-\lambda}Y^{-\kappa}
F_{\mu\nu\lambda\kappa}(X)
\nonumber \\ {}& & \qquad
{}+ \alpha' (R^{(2)})' Y^{+\mu} Y^{-\nu}  W_{\mu\nu}(X)
+ \alpha'^2 (e^\gamma R^{(2)} R^{(2)})' C(X) \biggr] \Biggl\} {,}
\label{LL}
\end{eqnarray}
As a result of calculations the parameters $\Lambda $,
$\bar \Lambda $ disappear from the
quantum algebra in this approximation. It means that any ordering with
various $\Lambda $, $\bar \Lambda $ corresponds
to the same quantum theories just as it is
in the case of free string where the only relevant ordering parameters are
$\alpha(0)$ and $\beta(0)$. This is an effect of linear
 approximation and there should appear  dependence on $\Lambda $, $\bar
\Lambda$ at higher levels. The regularization parameter $\epsilon$ has
cancelled out of the commutators too that is the quantum algebra
appeared in the form as if it were finite and
did not require any regularization procedure.

Eqs.(\ref{LL}) define the explicit form of the functions
$A_{\alpha\beta}$ (\ref{A}) in the theory under consideration.
Ghost contributions $G_{\alpha\beta}$ have the same structure as
in free string theory \cite{hwang} and cancel the $c-$valued
terms in (\ref{LL}) provided  that $D=26$ and
$\alpha(0)=\beta(0)=1$.

One can see from (\ref{LL}) that eqs.(\ref{a_0}) in our case
require $\gamma'(\tau,\sigma)=0$, $R^{(2)}{}'(\tau,\sigma)=0$.
Together with the condition (\ref{gamma}) it means that the
operators $L_n$, $\bar{L}_n$ form conformal algebra only
if $\gamma=const$ and so $R^{(2)} (\tau,\sigma)=0$. This does
not contradict the corresponding results of covariant
approaches. If effective equations of motion for background
fields are fulfilled the quantum effective action obtained by
means of functional integration over $X^\mu$ does not depend on
conformal factor $\gamma$.
If we write the metric in the form $g_{ab}=e^\gamma \bar{g}^{ab}$
then the effective action depends only on $\bar{g}^{ab}$.
But two independent components of $\bar{g}^{ab}$ are constrained
by gauge fixing functions. Therefore all physical results
(e.g. correlation functions of gauge invariant operators) will be the
same for any world sheet including the flat one.
So without any loss of generality one
can choose $\gamma$ to be constant and this is the choice that
is reproduced within our canonical formulation.

It should be noted that our approach is not
restricted by flat world sheets.  Quantum theory can be
formulated for any functions $\gamma$ but it is conformal
invariant only for constant $\gamma$.

If one chooses the conditions $\gamma=0$, $R^{(0)}=0$ then the
background fields $W(X)$ and $C(X)$ disappear from the classical
action (\ref{S}). Then to fulfill the eqs.(\ref{a_0}) the only
relevant background fields $Q(X)$ and $F(X)$ should obey the
following equations:
\begin{equation}
(\partial^2 + 4/\alpha')Q(X)=0, \qquad
(\partial^2 - 4/\alpha')F_{\mu\nu\lambda\kappa}(X)=0 ,
\label{shell}
\end{equation}
\begin{equation}
\partial^\mu F_{\mu\nu\lambda\kappa}(X)=0, \qquad
\partial^\lambda F_{\mu\nu\lambda\kappa}(X)=0,
\label{tranv}
\end{equation}
\begin{equation}
F^\mu{}_{\mu\lambda\kappa}(X)=0, \qquad
F_{\mu\nu}{}^\lambda{}_\lambda(X)=0.
\label{trace}
\end{equation}
The eqs.(\ref{shell}) are mass-shell conditions for background fields.
Other equations show that first massive level is described by a tensor
of fourth rank which is symmetrical and traceless in both pairs of
indices and transversal in all indices. This exactly corresponds
to the closed string spectrum \cite{green} and so our approach
gives the full set of correct linear equations for massive
background fields.

\section{Conclusion}

In our paper we proposed a general scheme allowing in
certain cases to construct guage invariant quantum formulation
starting with a classical theory without gauge invariance. In
usual quantization schemes classical first class constraints are
declared operators and they should obey some quantum gauge
algebra. Our prescription leads to more general scheme allowing to
build the operator of canonical BRST charge $\Omega$
generating quantum gauge transformations
for a non-gauge classical theory.

We applied the proposed prescription to the theory of closed
bosonic string coupled to massive background fields of the
first level which is not conformal invariant at the classical
level and found conditions of its quantum conformal invariance
in linear approximation.  Within this approach we derived the
full set of effective equations of motion for massive background
fields including the tracelessness conditions (\ref{trace})
deriving of which in covariant approaches represents
a problem~\cite{buchper,ellw}.

The obtained results show that our approach can be considered as
a general method allowing to construct gauge invariant
quantum theories starting from non-gauge classical models. In
particular we hope that this method provides a possibility for
deriving interacting effective equations of motion for massive
and massless background fields within the framework of canonical
formulation of string models and provides a justification of
covariant functional approach to string theory.

\section*{Acknowledgements}

The authors are grateful to E.S.~Fradkin, P.M.~Lavrov,
R.~Marnelius, B.~Ovrut, A.A.~Tseytlin and I.V.~Tyutin for
discussions of some aspects of the paper. The work was supported
by Russian Foundation for Basic Research, project No
96-02-16017.


\newpage

\begin{thebibliography}{50}
\bibitem{callan}
 E.S.Fradkin, A.A.Tseytlin, Phys.Lett. {\bf 158B}, 316 (1985),
 Nucl.Phys. {\bf 261B}, 1 (1985); \\
 C.G.Callan, O.Friedan, E.Martinec, M.J.Perry,
    Nucl.Phys. {\bf 262B}, 593 (1985);\\
 A. Sen, Phys.Rev.Lett. {\bf 55}, 1846 (1985);
     Phys.Rev. {\bf D32}, 2102 (1985).
\bibitem{tseyt}
 A.A.Tseytlin, Int.J.Mod.Phys. {\bf A4}, 1257 (1989);\\
  H.Osborn, Ann.Phys. (USA) {\bf 200}, 1 (1990).
\bibitem{Pol}
 A.M.Polyakov, Phys.Lett. {\bf B103}, 207 (1981).
\bibitem{buchshap}
 I.L.Buchbinder, I.L.Shapiro, S.G.Sibiryakov
   Nucl.Phys. {\bf 445B}, 109, (1995).
\bibitem{BFV}
 E.S.Fradkin, G.A.Vilkovisky, Phys.Lett. {\bf 55B}, 224 (1975);\\
 I.A.Batalin, G.A.Vilkovisky, Phys.Lett. {\bf 69B}, 409 (1977);\\
 E.S.Fradkin, T.E.Fradkina, Phys.Lett. {\bf 72B}, 343 (1978).
\bibitem{BatFr}
 I.A.Batalin, E.S.Fradkin, Phys.Lett. {\bf 128B}, 303 (1983);
  J.Math.Phys {\bf 25}, 2426 (1984);
  Riv.Nuovo Cim. {\bf 9}, 1 (1986);
  Phys.Lett. {\bf 180B}, 157 (1986);
  Nucl.Phys. {\bf 179B}, 514 (1987);
  Ann.Inst.H.Poincare, {\bf 49}, 145 (1988).
\bibitem{Hen}
 M.Henneaux, Phys.Rept. {\bf 126}, 1 (1985);\\
 M.Henneaux, C.Teitelboim. Quantization of Gauge Systems.
  Princeton Univ.Press., Princeton, N.J., 1992.
\bibitem{ovrut}
 J.M.F. Labastida, M.A.H. Vozmediano,
 Nucl.Phys. {\bf 312B}, 308 (1989);\\
 J.-C. Lee, B.A. Ovrut, Nucl.Phys. {\bf 336B}, 222 (1990).
\bibitem{buchper}
 I.L.Buchbinder, E.S.Fradkin, S.L.Lyakhovich, V.D.Pershin,
   Phys.Lett. {\bf 304B}, 239 (1993);
\bibitem{krykh}
I.L. Buchbinder, V.A.Krykhtin, V.D.Pershin, Phys.Lett. {\bf 348B},
 63 (1995); Phys.Atom.Nucl. (Yad.Fiz.) {\bf 59}, 332 (1996).
\bibitem{sathiap}
B. Sathiapalan, Nucl.Phys. {\bf 326B}, 376 (1989),
{\bf 415B}, 332 (1994); Mod.Phys.Lett. {\bf A11}, 571 (1996).
\bibitem{ellw}
 V. Ellwanger, J. Fuchs, Nucl.Phys. {\bf 312B}, 95 (1989);\\
 J. Hughes, J. Liu, J. Polchinski, Nucl.Phys. {\bf 316B}, 15
 (1989).
\bibitem{hwang}
S. Hwang, Phys.Rev. {\bf D28}, 2614 (1983).
R. Marnelius, Nucl.Phys. {\bf 211B}, 14 (1984); {\bf 221B}, 409
(1983).
\bibitem{fujiw}
T. Fujiwara, Y. Igarashi, J. Kubo, K. Maeda, Nucl.Phys. {\bf 391B},
211 (1993);\\
T. Fujiwara, Y. Igarashi, M. Koseki, R. Kuruki, T. Tabei,
Nucl.Phys.  {\bf 425B}, 289 (1994); \\
\bibitem{buch91}
 I.L.Buchbinder, E.S.Fradkin, S.L.Lyakhovich, V.D.Pershin, Int.J.Mod.Phys.
 {\bf 6A}, 1211 (1991).
\bibitem{buch95}
 I.L.Buchbinder, B.R.Mistchuk, V.D.Pershin, Phys.Lett. {\bf 353B}, 257
 (1995).
\bibitem{Fubini}
 S.Fubini, M.Roncadelli, Phys.Lett. {\bf 203B}, 433 (1988).
\bibitem{Das}
 S.R.Das, B.Sathiapalan, Phys.Rev.Lett. {\bf 56}, 2654 (1986);\\
 C.Itoi, Y.Watabiki, Phys.Lett. {\bf 198B}, 486 (1987);\\
 A.A.Tseytlin, Phys.Lett. {\bf 264B}, 311 (1991).
\bibitem{berezin}
 F.A.Berezin, M.A.Shubin, Schr\"odinger Equation,
   Moscow Univ.Press, Moscow, 1983 (in Russian).
\bibitem{green}
 M.B.Green, J.H.Schwarz, E.Witten, Superstring Theory,
   Cambridge Univ.Press, Cambridge, 1987.
\end{thebibliography}
\end{document}